\documentclass[a4paper,twoside,12pt]{article}
\usepackage{graphicx}
\usepackage{bm}

\usepackage[left=15mm,right=15mm,top=20mm,bottom=30mm,bindingoffset=1.5cm]{geometry}

\usepackage{indentfirst}           % abzats
\usepackage{epstopdf}
\usepackage{fancyhdr}
\usepackage{amsmath}
\usepackage{amssymb}
\usepackage{tikz}
\usepackage{multirow}
\usetikzlibrary{arrows}
\usetikzlibrary{decorations.pathmorphing}
\usetikzlibrary{decorations.markings}
\usetikzlibrary{calc}

\tikzset{
  >=stealth',
  midarrow/.style={
    postaction={
      decorate,
      decoration={markings, mark=at position .55 with {\arrow{>}}}
    }
  },
  fermion/.style=midarrow,
  photon/.style={
    decorate,
    decoration={snake, amplitude=2pt, segment length=8pt}
  },
  boson/.style={
    decorate,
    decoration={snake, amplitude=2pt, segment length=8pt}
  },
  gluon/.style={
    decorate,
    decoration={coil, amplitude=4pt, segment length=5pt}
  },
  scalar/.style=densely dashed,
  arrowsnake/.style={
    preaction={photon, draw},
    postaction=midarrow
  }
}

\def\Journal#1#2#3#4{{#1} {\bf #2}, #3 (#4)}

\def\PLB{{\em Phys. Lett.}  B}

\def\EPJ{{\em Eur. Phys. J.} C}
\def\CPC{\em Comm. Phys. Comm.}

\def\PR{\em Phys. Rep.}

\title{ \bf LHC as a photon-photon collider}

\author{M.I. Vysotsky$^{1,2}$ and E.V. Zhemchugov$^{1,2}$}
\date{}
\begin{document}

\maketitle
\begin{center}
{\em 
$^1$ Institute for Theoretical and Experimental Physics, 117218, Moscow, Russia, \\
$^2$ Moscow Engineering Physics Institute, 115409, Moscow, Russia\\ 
}
\end{center}

\vspace{5mm}

\begin{abstract}
Equivalent photon approximation is used to
calculate fiducial cross sections for dimuon production in
ultraperipheral proton-proton and lead-lead collisions. Analytical
formulae taking into account experimental cuts are derived. The
results are compared with the measurements reported by the ATLAS
collaboration.
\end{abstract}
\medskip

This contribution is based on the paper by M.I. Vysotsky and E.V.
Zhemchugov \cite{1}.

  \begin{center}
    \hfill
    \includegraphics[height=0.1\textheight]{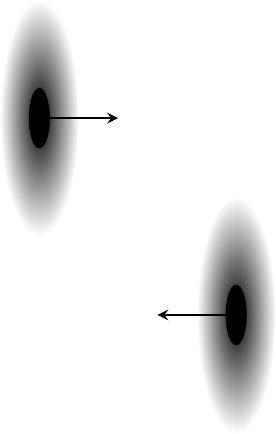}
    \hfill
    \includegraphics[height=0.1\textheight]{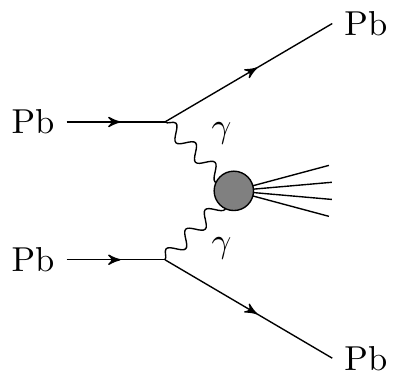}
    \hfill
    \phantom{.}
    \\
    $\sigma \sim Z^4$
  \end{center}
\bigskip

Integrated luminosity in $pp$ collisions at 13~TeV, provided by
the LHC to CMS in Run~2: 119~fb$^{-1}$ current, 150~fb$^{-1}$
expected. Integrated luminosity in Pb~Pb collisions provided by
the LHC to CMS in HI~run (2015): $0.7~\text{nb}^{-1}$. Luminosity
ratio (expected): $2.1 \cdot 10^8$. For Pb, $Z = 82$; $Z^4 \approx
4.5 \cdot 10^7$. There could be about 5 times more events of New
Physics in $\gamma \gamma$ collisions in $pp$  than there were in
Pb~Pb data. HI~run duration was $\approx 20$~days, Run~2 duration
was $\approx 500$~days (not counting the 2015). A question on
whether the heavy ion luminosity should be increased might open to
discussion.

Our goal is to derive analytical formulas to describe experimental
data.

Experimental data:
  \begin{enumerate}
    \item ATLAS \cite{2}: $pp \to pp \mu^+ \mu^-$ at collision
    energy 13~TeV with integrated luminosity of~$3.2~\text{fb}^{-1}$. \\
    Fiducial cross section: $3.12 \pm 0.07~\text{(stat.)} \pm
    0.10~\text{(syst.)}$~pb. \\
    Cuts on the muon system parameters: \\
    \begin{tabular}{|ccc|}
      \hline
      Invariant mass range & Transverse momentum & Pseudorapidity \\
      $12~\text{GeV} < m_{\mu \mu} < 30~\text{GeV}$
      & $> 6~\text{GeV}$
      & $< 2.4$
      \\
      $30~\text{GeV} < m_{\mu \mu} < 70~\text{GeV}$
      & $> 10~\text{GeV}$
      & $< 2.4$
      \\ \hline
    \end{tabular}
    \item ATLAS \cite{3}: $\text{Pb} \; \text{Pb} \to \text{Pb} \;
    \text{Pb} \; \mu^+ \mu^-$ at collision energy per nucleon pair $5.02$~TeV with
    integrated luminosity of $515~\mu\text{b}^{-1}$.\\
    Fiducial cross section:
    $32.2 \pm 0.3~\text{(stat.)}^{+4.0}_{-3.4}~\text{(syst.)}~\mu\text{b}$. \\
    Cuts on the muon system parameters: \\
    \begin{tabular}{|ccc|}
      \hline
      Invariant mass range & Transverse momentum & Pseudorapidity \\
      $10~\text{GeV} < m_{\mu \mu} < 100~\text{GeV}$
      & $> 4~\text{GeV}$
      & $< 2.4$
      \\ \hline
    \end{tabular}
  \end{enumerate}

Equivalent photon approximation:

E.~Fermi(1924), C.~F.~V.~Weizs\"acker (1934), E.~J.~Williams
(1935), L.~D.~Landau, E.~M.~Lifshitz (1934).

  \begin{tikzpicture}
    \coordinate (O1) at (-3, 0);
    \coordinate (O2) at ( 3, 0);

    \path [shade, inner color=black, outer color=white] (O1) circle (1);
    \draw [fill] (O1) circle (0.25);

    \path [shade, inner color=black, outer color=white]
      (O2) ellipse (0.4 and 1.2);
    \draw [fill]
      (O2) ellipse (0.1 and 0.3);

    \draw [->, >=stealth] (-1, 0) -- node [below] {$\gamma \gg 1$} (2, 0);

    \draw [photon]
      (O1) ++(0.4, 0.4) -- ++(25:1) node [right] {$(0, q_x, q_y, q_z)$};
    \draw [photon]
      (O2) ++(0.1, 0.4)
      -- ++(0:1) node [right] {$(\sqrt{\gamma^2 - 1} q_z, q_x, q_y, \gamma q_z)$};
  \end{tikzpicture}

  Photon virtuality: $-q^2 = q_x^2 + q_y^2 + q_z^2 \ll (\gamma q_z)^2 \equiv
  \omega^2$: so, (almost) real photons!

From the textbook \cite{x} and the review paper \cite{y} we have:

\begin{equation}
\begin{array}{ll}
     n(\vec q)
    &= \frac{Z^2 \alpha}{\pi^2}
       \frac{\vec q_\perp^{\, 2}}{\omega q^4}
     = \frac{Z^2 \alpha}{\pi^2}
       \frac{\vec q_\perp^{\, 2}}
            {\omega\left( q_\perp^{\, 2} + (\omega / \gamma)^2 \right)^2},
\end{array}
\label{1}
\end{equation}
\begin{equation}
\begin{array}{ll}
    n(\omega)
    &= \int n(\vec q) \mathrm{d}^2 q_\perp= 2 \pi
       \int\limits_0^{\hat q}
         n(\vec q) q_\perp
       \mathrm{d} q_\perp =
    \\   & = \frac{Z^2 \alpha}{\pi \omega}
       \left\{
           \ln \left[ 1 + \left( \frac{\hat q \gamma}{\omega} \right)^2 \right]
         - \frac{1}{1 + \left( \omega/(\hat q \gamma) \right)^2}
       \right\}
    \approx (\omega \ll \hat q \gamma) \approx
     \frac{2 Z^2 \alpha}{\pi \omega} \ln \frac{\hat q \gamma}{\omega}  \\
    ~&\hat q = \; ?
\end{array}
  \label{2}
\end{equation}

For the proton, $\hat q \approx \Lambda_\text{QCD} = 0.2$--$0.3$~GeV.

  Dirac form factor:
  \begin{gather*}
    \mathcal{J}_\mu = F(q^2) \bar \psi \gamma_\mu \psi, \
    F(q^2) \approx \frac{1}{\left(1 - \frac{q^2}{\Lambda^2} \right)^2} \ \
    (-q^2 \ll 4 m_p^2), \
    \Lambda^2 = 0.71~\text{GeV}^2.
  \end{gather*}

  EPA spectrum with form factor:
  \begin{align*}
    n'(\vec q)
    &= \frac{Z^2 \alpha}{\pi^2}
       \frac{\vec q_\perp^{\, 2}}{\omega q^4}
       \frac{1}{\left( 1 - \frac{q^2}{\Lambda^2} \right)^2},
    \\
    n'(\omega)
    &= \int n'(\vec q) \mathrm{d}^2 q
     = 2 \pi \int\limits_0^\infty n'(\vec q) q_\perp \mathrm{d} q_\perp
%    &= \frac{Z^2 \alpha}{\pi \omega}
%       \left.
%         \left[ (4 a + 1) \ln \left( 1 + \frac{1}{a} \right)
%         - \frac{24 a^2 + 42 a + 17}{6 (a + 1)^2}
%       \right\rvert_{a = \left( \frac{\omega}{\Lambda \gamma} \right)^2}
%    \\
%    &\xrightarrow[a \to 0]{}
     \approx
     \frac{2 Z^2 \alpha}{\pi \omega}
     \left( \ln \frac{\Lambda \gamma}{\omega} - \frac{17}{12} \right)
     \ (\omega \ll \Lambda \gamma)\; .
  \end{align*}
In the leading logarithmic approximation $n'(\omega)\approx
n(\omega)$, hence $\hat q = \Lambda \mathrm{e}^{-\frac{17}{12}}
\approx 204$~MeV. For Pb, $\Lambda \approx
80$~MeV~[hep-ph/0606069] and $\hat q \approx 20$~MeV.

An integral over $q_\perp$ should be cut at min($\hat q,
m_{\mu\mu}) = \hat q$.

\bigskip

Total cross section:

\small

 \begin{figure}[!tbh]
  \centering
  \begin{tikzpicture}
    % P1i -- A ---- P1o
    %         \
    %          B -- M1
    %          |
    %          C -- M2
    %         /
    % P2i -- D ---- P2o

    \coordinate (A)   at (-0.7,  1.2);
    \coordinate (B)   at (   0,  0.5);
    \coordinate (C)   at (   0, -0.5);
    \coordinate (D)   at (-0.7, -1.2);
    \coordinate (P1i) at (-1.2,  1.2);
    \coordinate (P1o) at (   1,  1.2);
    \coordinate (P2i) at (-1.2, -1.2);
    \coordinate (P2o) at (   1, -1.2);
    \coordinate (M1)  at (   1,  0.5);
    \coordinate (M2)  at (   1, -0.5);

    \draw[fermion] (P1i) node [left] {$p$} -- (A);
    \draw[fermion] (A)   -- (P1o) node [right] {$p$};
    \draw[fermion] (P2i) node [left] {$p$} -- (D);
    \draw[fermion] (D)   -- (P2o) node [right] {$p$};
    \draw[photon]  (A)   -- (B);
    \draw[photon]  (D)   -- (C);
    \draw[fermion] (M1)  node [right] {$\mu$} -- (B);
    \draw[fermion] (B)   -- (C);
    \draw[fermion] (C)   -- (M2) node [right] {$\mu$};

    \draw [->] ([yshift=-2mm]$(A)!0.1!(B)$)
               to node [midway, below left] {$(\omega_1, \vec q_1)$}
               ([yshift=-2mm]$(A)!0.6!(B)$);
    \draw [->] ([yshift=2mm]$(D)!0.2!(C)$)
               to node [midway, above left] {$(\omega_2, \vec q_2)$}
               ([yshift=2mm]$(D)!0.7!(C)$);
  \end{tikzpicture}
  ~
  \begin{tikzpicture}
    % Cross diagram
    \coordinate (A)   at (-0.7,  1.2);
    \coordinate (B)   at (   0,  0.5);
    \coordinate (C)   at (   0, -0.5);
    \coordinate (D)   at (-0.7, -1.2);
    \coordinate (P1i) at (-1.2,  1.2);
    \coordinate (P1o) at (   1,  1.2);
    \coordinate (P2i) at (-1.2, -1.2);
    \coordinate (P2o) at (   1, -1.2);
    \coordinate (M1)  at (   1,  0.5);
    \coordinate (M2)  at (   1, -0.5);

    \draw[fermion] (P1i) node [left] {$p$} -- (A);
    \draw[fermion] (A)   -- (P1o) node [right] {$p$};
    \draw[fermion] (P2i) node [left] {$p$} -- (D);
    \draw[fermion] (D)   -- (P2o) node [right] {$p$};
    \draw[photon]  (A)   -- (C);
    \draw[photon]  (D)   -- (B);
    \draw[fermion] (M1)  node [right] {$\mu$} -- (B);
    \draw[fermion] (B)   -- (C);
    \draw[fermion] (C)   -- (M2) node [right] {$\mu$};

    \draw [->] ([xshift=-2mm]$(A)!0.2!(C)$)
               to node [midway, left] {$(\omega_1, \vec q_1)$}
               ([xshift=-2mm]$(A)!0.5!(C)$);
    \draw [->] ([xshift=-2mm]$(D)!0.2!(B)$)
               to node [midway, left] {$(\omega_2, \vec q_2)$}
               ([xshift=-2mm]$(D)!0.5!(B)$);

  \end{tikzpicture}

\end{figure}
  \hfill \phantom{.}
  \begin{equation*}
    \mathrm{d} \sigma(pp(\gamma \gamma) \to \mu^+ \mu^- pp)
     = \sigma(\gamma \gamma \to \mu^+ \mu^-)
       \cdot
       n(\omega) n(\omega') \mathrm{d} \omega \mathrm{d} \omega',
  \end{equation*}

  \resizebox{\hsize}{!}{$
     \sigma(\gamma \gamma \to \mu^+ \mu^-)
      = \frac{4 \pi \alpha^2}{s}
        \left[
          \left( 1 + \frac{4 m_\mu^2}{s} - \frac{8 m_\mu^4}{s^2} \right)
          \ln \frac{1 + \sqrt{1 - \frac{4 m_\mu^2}{s}}}
                   {1 - \sqrt{1 - \frac{4 m_\mu^2}{s}}}
          - \left( 1 + \frac{4 m_\mu^2}{s} \right)
            \sqrt{1 - \frac{4 m_\mu^2}{s}}
        \right]
  $}

  Integration constraints:
  $s \equiv (q + q')^2 > 4 m_\mu^2$, $\omega < \hat q \gamma$, $\omega' < \hat q \gamma$.
  \\
  Change of variables: $s = 4 \omega \omega'$, $x = \frac{\omega}{\omega'}$;
  $\omega = \sqrt{\frac{s x}{4}}$, $\omega' = \sqrt{\frac{s}{4 x}}$
  $(2 m_\mu)^2 < s < (2 \hat q \gamma)^2$,
  $\frac{(2 \hat q \gamma)^2}{s} > x > \frac{s}{(2 \hat q \gamma)^2}$.

    \begin{equation*}
    \sigma(pp(\gamma \gamma) \to pp \mu^+ \mu^-)
    \approx
    8 \cdot \frac{28}{27} \frac{\alpha^4}{\pi m_\mu^2}
    \ln^3 \frac{\hat q \gamma}{m_\mu}
    \approx
    2.2 \cdot 10^5~\text{pb}\; .
  \end{equation*}

Cut on invariant mass: $70~\text{GeV} > \sqrt{s} > 12~\text{GeV}
\gg m_\mu = 0.106~\text{GeV}$.
$$
\sigma(\gamma \gamma \to \mu^+ \mu^-)
    \approx \dfrac{4 \pi \alpha^2}{s} \left( \ln \dfrac{s}{m_\mu^2} - 1 \right)
    $$

  \begin{gather*}
    \sigma^{(\hat s)}_\text{fid.}
      (pp(\gamma \gamma) \to pp \mu^+ \mu^-)
     = \int\limits_{\hat s_\text{min}}^{\hat s_\text{max}}
         \mathrm{d} s \,
         \sigma(\gamma \gamma \to \mu^+ \mu^-)
         \int\limits_{s / (2 \hat q \gamma)^2}^{(2 \hat q \gamma)^2 / s}
           \dfrac{\mathrm{d} x}{8 x}
           n \left( \sqrt{\dfrac{s x}{4}} \right)
           n \left( \sqrt{\dfrac{s}{4 x}} \right)=
    \\
     = \left.
         \frac{8 \alpha^4}{3 \pi}
         \frac{1}{(2 \hat q \gamma)^2}
         \frac{1}{z}
         \left[
           \ln^4 z
           + \left( 2 \ln \frac{2 \hat q \gamma}{m_\mu} + 3 \right)
             \left( \ln^3 z + 3 \ln^2 z + 6 \ln z + 6 \right)
         \right]
       \right\rvert_{z = \frac{\hat s_\text{min}}{(2 \hat q \gamma)^2}}
                   ^{\frac{\hat s_\text{max}}{(2 \hat q \gamma)^2}}
\approx 59.6~\text{pb}.
  \end{gather*}

Cuts on transverse momentum and on pseudorapidity were also
imposed.

Analogous formulas with evident substitutions were used in
\cite{1} to describe $\text{Pb} \; \text{Pb} \; (\gamma \gamma)
\to \text{Pb} \; \text{Pb} \; \mu^+ \mu^-$ reaction.

\begin{center}

$pp(\gamma \gamma) \to \mu^+ \mu^- pp$

  \begin{tabular}{|lcc|}
    \hline
    \multicolumn{2}{|l}{No cuts} & $2.2 \cdot 10^5$~pb \\ \hline
    $12~\text{GeV} < \sqrt{s} < 30~\text{GeV}$
    & $54.0$~pb
    & \multirow{2}{*}{$59.6$~pb}
    \\
    $30~\text{GeV} < \sqrt{s} < 70~\text{GeV}$
    & $5.65$~pb
    &
    \\ \hline
    $12~\text{GeV} < \sqrt{s} < 30~\text{GeV}$,
    $p_T > 6~\text{GeV}$
    & $5.37$~pb
    & \multirow{2}{*}{$6.28$~pb}
    \\
    $30~\text{GeV} < \sqrt{s} < 70~\text{GeV}$,
    $p_T > 10~\text{GeV}$
    & $0.91$~pb
    &\\ \hline
    $12~\text{GeV} < \sqrt{s} < 30~\text{GeV}$,
    $p_T > 6~\text{GeV}$,
    $|\eta | < 2.4$
    & $2.85$~pb
    & \multirow{2}{*}{$3.35$~pb}
    \\
    $30~\text{GeV} < \sqrt{s} < 70~\text{GeV}$,
    $p_T > 10~\text{GeV}$,
    $|\eta| < 2.4$
    & $0.50$~pb
    &\\ \hline
  \end{tabular}
  \vfill
  $\text{Pb} \; \text{Pb} \; (\gamma \gamma)
  \to \mu^+ \mu^- \text{Pb} \; \text{Pb}$

  \begin{tabular}{|lc|}
    \hline
    No cuts & $2.80 \cdot 10^6~\mu\text{b}$ \\
    $10~\text{GeV} < \sqrt{s} < 100~\text{GeV}$ & $119~\mu\text{b}$ \\
    also $p_T > 4$~GeV & $34.2~\mu\text{b}$ \\
    also $|\eta| < 2.4$ & $30.9~\mu\text{b}$ \\ \hline
  \end{tabular}

  \end{center}

\bigskip

Theoretical predictions obtained with the help of Monte Carlo
method:
\begin{enumerate}
\item the {\tt SuperCHIC} program \cite{4} gives
$$
  \sigma_\text{fid.} (\text{$pp$} \; (\gamma \gamma) \to \text{$pp$} \; \mu^+ \mu^-)= 3.45 \pm 0.05~\text{pb};
$$
\item EPA prediction corrected for absorptive effects \cite{5}
gives
$$
  \sigma_\text{fid.}(\text{$pp$} \; (\gamma \gamma) \to \text{$pp$} \; \mu^+ \mu^-) = 3.06 \pm 0.05~\text{pb}.
$$
\item calculations with the help of the STARLIGHT program
\cite{6}:
$$
  \sigma_\text{fid.}
        (\text{Pb} \; \text{Pb} \; (\gamma \gamma) \to \text{Pb} \; \text{Pb} \; \mu^+ \mu^-)
  = 31.64 \pm 0.04~\mu\text{b}.
$$
\end{enumerate}

%\begin{figure}[!tbh]
\begin{figure}
  %\centering
  \includegraphics[height=0.2\textheight]{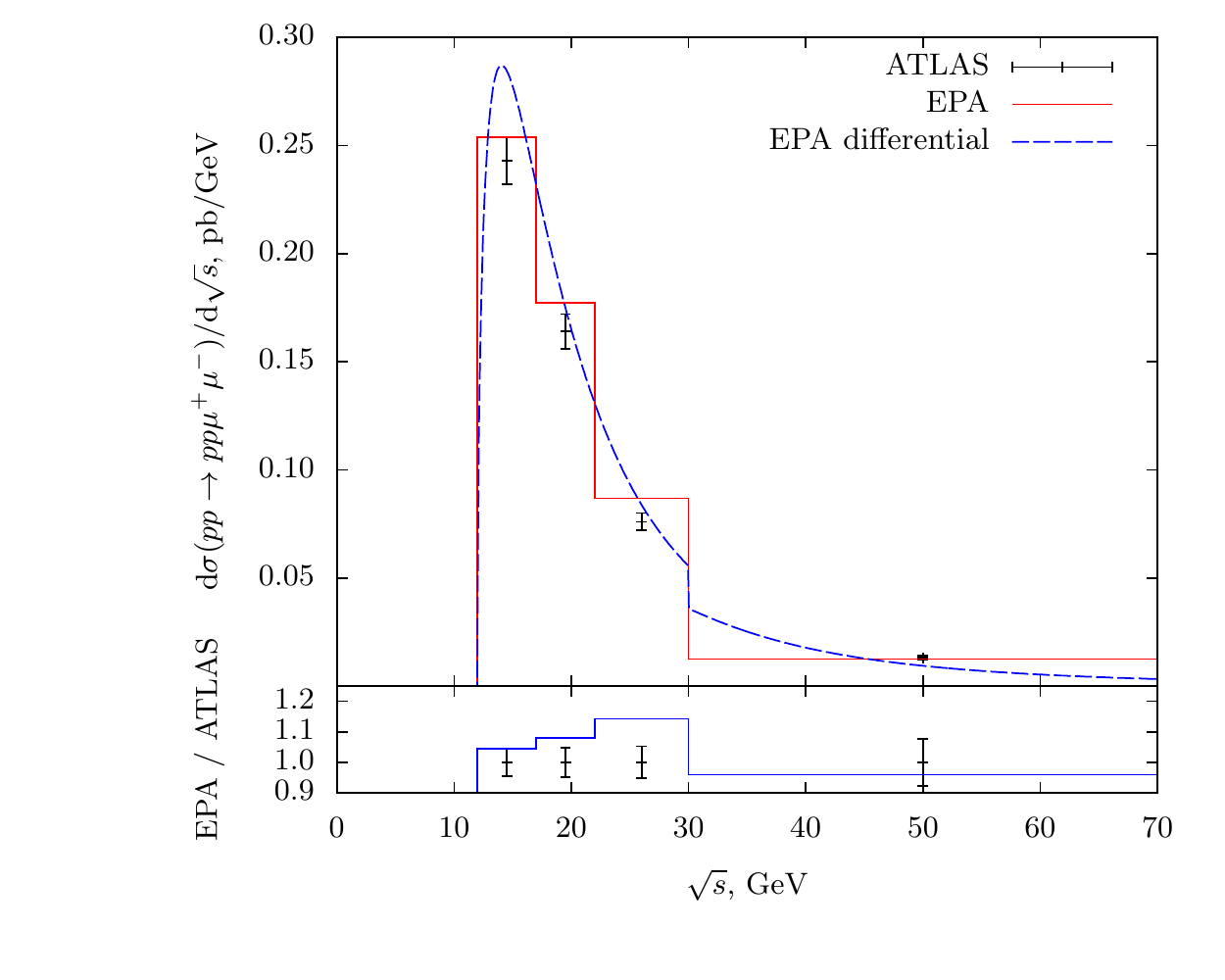}
  \hfill
    \includegraphics[height=0.2\textheight]{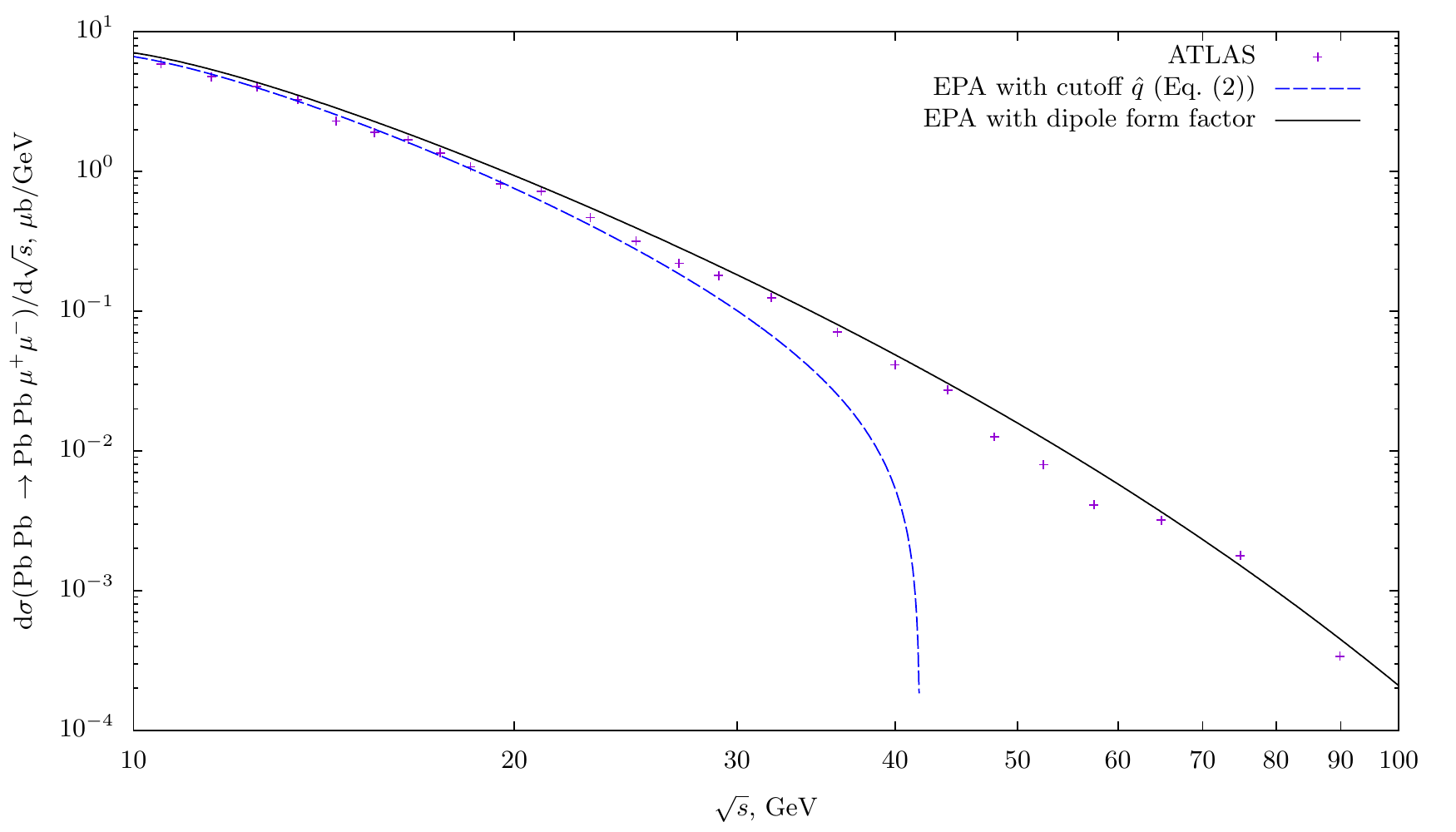}
    \hfill
    \phantom{.}
  \caption{\emph{Left plot:} fiducial cross section for the $pp (\gamma \gamma)
    \to pp \mu^+ \mu^-$ reaction at proton-proton collision energy 13~TeV
    with the cuts described in the first table. Points are the
    experimental data presented in Table~3 of$~\cite{2}$. The dashed
    line is the differential cross section calculated by us.
    The histogram is the differential cross section
    integrated according to the bins also presented in Table~3 of$~\cite{2}$.
  \emph{Lower plot:} ratio of the calculated cross
    section to the experimental points.
{\emph{Right plot:} fiducial cross section for the $\text{Pb} \;
\text{Pb} \; (\gamma\gamma) \to \text{Pb} \; \text{Pb} \; \mu^+
\mu^-$ reaction at collision energy per nucleon pair $5.02$~TeV
with the experimental cuts described in the second table. Points
are experimental data from the left plot of Fig.~3 of$~\cite{3}$.
The dashed line is calculated using the EPA spectrum cut at
$\hat{q} = 20$ MeV. The solid line is calculated using the
spectrum with dipole form factor, and the form factor parameter
$\Lambda = 80$ MeV. }}
\end{figure}

We have derived analytical formulas for the fiducial cross section
of lepton pair production in peripheral collisions of charged
particles. Experimental data were found to be in agreement with
fiducial cross sections calculated for the reactions $pp(\gamma
\gamma) \to pp \mu^+ \mu^-$ and $\text{Pb} \; \text{Pb} \; (\gamma
\gamma) \to \text{Pb} \; \text{Pb} \; \mu^+ \mu^-$.

\bigskip

This is the written version of the talk presented by M.~Vysotsky at Windows on
the Universe 2018, Rencontres du Vietnam, Quy Nhon, Vietnam, August 5--11, 2018,
and the 20th Annual RDMS CMS Collaboration Conference, Tashkent, Uzbekistan,
September 12--15, 2018.

\bigskip

We were supported by the RFBR grant 16-02-00342.

\end{document}